\documentclass[aoas,MSNbibl,nameyear,dvips]{arximspdf}
\usepackage{stfloats}
\usepackage{graphicx}

\doi{10.1214/13-AOAS680} 
\volume{8}
\issue{1}
\pubyear{2014}
\firstpage{259}
\lastpage{285}

\makeatletter
  \fnbelowfloat

\newcommand{\rrVert}{\Vert}
\newcommand{\llVert}{\Vert}
\renewcommand{\citep}[1]{(\citeauthor{#1} \citeyear{#1})}
\newcommand{\eqref}[1]{(\ref{#1})}
\def\cal{\mathcal}

\newcommand{\argmin}{\mathop{\operatorname{argmin}}}

\def\M{{\cal M}}\def\S{{\cal S}}
\def\X{{\cal X}}
\def\cov{\operatorname{Cov}}

\def\Iter{\mathsf{Iter}}

\newproclaim{rmk}{Remark}
\newproclaim{eg}{Example}
\newproclaim{defn}{Definition}

\makeatother

\begin{document}
\begin{frontmatter}

\title{$\gamma$-SUP: A clustering algorithm for cryo-electron microscopy images of asymmetric particles}
\runtitle{$\gamma$-SUP}

\begin{aug}
\author[A]{\fnms{Ting-Li} \snm{Chen}\thanksref{t1}\ead[label=e1]{tlchen@stat.sinica.edu.tw}},
\author[A]{\fnms{Dai-Ni} \snm{Hsieh}\thanksref{t1}\ead[label=e2]{dnhsieh@stat.sinica.edu.tw}},
\author[C]{\fnms{Hung} \snm{Hung}\thanksref{t2}\ead[label=e3]{hhung@ntu.edu.tw}},
\author[A]{\fnms{I-Ping} \snm{Tu}\corref{}\ead[label=e4]{iping@stat.sinica.edu.tw}\thanksref{t1}},
\author[E]{\fnms{Pei-Shien} \snm{Wu}\thanksref{t3}\ead[label=e5]{pei.shien.wu@duke.edu}},
\author[A]{\fnms{Yi-Ming} \snm{Wu}\thanksref{t1}\ead[label=e6]{marinesean@gmail.com}},
\author[A]{\fnms{Wei-Hau} \snm{Chang}\ead[label=e7]{weihau@chem.sinica.edu.tw}\thanksref{t1}}
\and
\author[A]{\fnms{Su-Yun} \snm{Huang}\thanksref{t1}\ead[label=e8]{syhuang@stat.sinica.edu.tw}}
\runauthor{T.-L. Chen et al.}
\affiliation{Academia Sinica\thanksmark{t1},
National Taiwan University\thanksmark{t2} and
Duke University\thanksmark{t3}}
\address[A]{T.-L. Chen\\
D.-N. Hsieh\\
I-P. Tu\\
Y.-M. Wu\\
W.-H. Chang\\
S.-Y. Huang\\
Institute of Statistical Science\\
Academia Sinica\\
Taipei, Taiwan 11529\\
\printead{e1}\\
\phantom{E-mail:\ }\printead*{e2}\\
\phantom{E-mail:\ }\printead*{e4}\\
\phantom{E-mail:\ }\printead*{e6}\\
\phantom{E-mail:\ }\printead*{e7}\\
\phantom{E-mail:\ }\printead*{e8}} 
\address[C]{H. Hung\\
Institute of Epidemiology\\
\quad and Preventive Medicine\\
National Taiwan University\\
Taipei, Taiwan 10055\\
\printead{e3}}
\address[E]{P.-S. Wu\\
Department of Biostatistics\\
\quad and Bioinformatics\\
Duke University\\
Durham, North Carolina 27710\\
USA\\
\printead{e5}}
\end{aug}

\received{\smonth{9} \syear{2012}}
\revised{\smonth{8} \syear{2013}}

%
\begin{abstract}
Cryo-electron microscopy (cryo-EM) has recently emerged as a powerful
tool for obtaining three-dimensional (3D) structures of biological
macromolecules in native states. A minimum cryo-EM image data set for
deriving a meaningful reconstruction is comprised of thousands of
randomly \mbox{orientated} projections of identical particles photographed
with a small number of electrons. The computation of
3D structure from 2D projections requires clustering, which aims to
enhance the signal to noise ratio in each view by grouping similarly
oriented images. Nevertheless, the prevailing clustering techniques are
often compromised by three characteristics of cryo-EM data: high noise
content, high dimensionality and large number of clusters. Moreover,
since clustering requires registering images of similar orientation
into the same pixel coordinates by 2D alignment, it is desired that the
clustering algorithm can label misaligned images as outliers. Herein,
we introduce a clustering algorithm $\gamma$-SUP to model the data with
a $q$-Gaussian mixture and adopt the minimum $\gamma$-divergence for
estimation, and then use a self-updating procedure to obtain the
numerical solution. We apply $\gamma$-SUP to the cryo-EM images of two
benchmark macromolecules, RNA polymerase II and ribosome. In the former
case, simulated images were chosen to decouple clustering from
alignment to demonstrate $\gamma$-SUP is more robust to misalignment
outliers than the existing clustering methods used in the cryo-EM
community. In the latter case, the clustering of real cryo-EM data by
our $\gamma$-SUP method eliminates noise in many views to reveal true
structure features of ribosome at the projection level.
\end{abstract}

%

\end{frontmatter}

\section{Introduction and motivating example}

Determining 3D atomic structure of large biological molecules is
important for elucidating the physicochemical mechanisms underlying
vital processes. In 2006, the Nobel Prize in chemistry was awarded to
structural biologist Roger D. Kornberg for his studies of the molecular
basis of eukaryotic transcription, in which he obtains for the first
time an actual picture at the molecular level of the structure of RNA
polymerase II during the stage of actively making messenger RNA. Three
years later in 2009, the same prize went to three X-ray
crystallographers, Venkatraman Ramakrishnan, Thomas~A. Steitz and
Ada~E. Yonath, for their revelation at the atomic level of the
structure and workings of the ribosome, with its even larger and more
complex machinery that translates the information contained in RNA into
a poly-peptide chain. Despite these successes, most large proteins have
resisted all attempts at crystallization. This has led to the emergence
of cryo-electron microscopy (cryo-EM), an alternative to \mbox{X-ray}
crystallography for obtaining 3D structures of macromolecules, since it
can focus electrons to form images without the need of crystals
[\citet{Henderson,van,Saibil}, \citeauthor{Frank1}
(\citeyear{Frank1,Frank2,Frank3}, \citet{Jiang,Zhou,Grassucci}].

%
\begin{figure}

\includegraphics{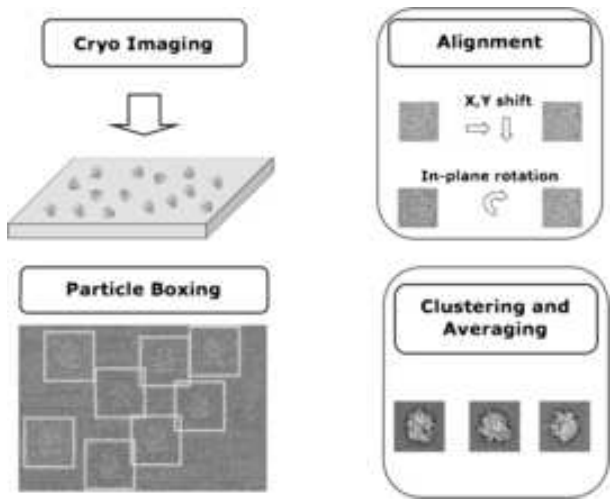}

\caption{The flowchart for cryo-EM analysis.}
\label{flowChartNew}
\end{figure}

A flowchart for cryo-EM analysis is shown in Figure~\ref{flowChartNew}.
In the sample preparation step, macromolecules obtained through
biochemical purification in the native condition are frozen so rapidly
that the surrounding water forms a thin layer of amorphous ice to embed
the molecules [\citet{Lepault,Adrian,Dubochet}]. The data obtained by
cryo-EM imaging consists of a large number of particle images
representing projections from random orientations of the macromolecule.
An essential step in reconstructing the 3D structure from these images
is to determine the 3D angular relationship of the 2D projections,
which is a challenging task for the following reasons. First, the
images are of poor quality, as they are heavily contaminated by shot
noise induced by the extremely low number of electrons used to prevent
radiation damage. Second, in contrast to X-ray crystallography, which
restricts the conformation of the macromolecule in the crystal, cryo-EM
data retain any conformation variations of the macromolecule that exist
in their original solution state, which means the data is a mixture of
conformations on top of the viewing angles. The left panel of
Figure~\ref{flowChartNew} explains how the 2D cryo-EM images are
collected and the right panel shows the commonly used strategy to
improve the signal-to-noise ratio (SNR) of those images.

Given reasonably clear 2D projections, an ab initio approach based on
the projection-slice theorem for 3D reconstruction from 2D projections
is available [\citet{Bracewell}]. This theorem states that any two
nonparallel 2D projections of the same 3D object would share a common
line in Fourier space. This common-line principle underlies the first
3D reconstruction of a spherical virus with icosahedral symmetry from
electron microscope micrographs [\citet{Crowther2}]. The same principle
was further applied to the problem of angular reconstruction of the
asymmetric particle [\citet{van2}]. In practice, a satisfactory solution
depends on the quality of the images and becomes increasingly
unreachable for raw cryo-EM images as the SNR gets too low. It is
therefore necessary to enhance the SNR of each view by averaging many
well aligned cryo-EM images coming from a similar viewing angle.
Clustering is thus aimed at grouping together the cryo-EM images with
nearly the same viewing angle. This step requires pre-aligning the
images because incorrect in-plane rotations and shiftings would prevent
successful clustering. Failure to cluster images into homogeneous
groups would, in turn, render the determination of the 3D structure
unsatisfactory. Currently, most approaches for clustering cryo-EM
images are rooted in the $k$-means method, which has been found to be
unsatisfactory [\citet{Yang}].

Here, we focus on the clustering step and assume that the image
alignment has been carried through. In the vast number of clustering
algorithms developed, two major approaches are taken. A model-based
approach [\citet{Banfield}] models the data as a mixture of parametric
distributions and the mean estimates are taken to be the cluster
centers. A distance-based approach enforces some ``distance'' to
measure the similarity between data points, with notable examples being
hierarchical clustering [\citet{Hartigan}], the $k$-means algorithm
[\citet{McQueen,Lloyd}] and the SUP clustering algorithm [\citet{Chen,Shiu}]. In this paper, we combine these two approaches to propose a
clustering algorithm $\gamma$-SUP. We model the data with a
$q$-Gaussian mixture [\citet{Amari,Eguchi2}] and adopt the $\gamma
$-divergence [\citet{Fujisawa,Cichocki,Eguchi2}] for measuring the
similarity between the empirical distribution and the model
distribution, and then borrow the self-updating procedure from SUP
[\citet{Chen,Shiu}] to obtain a numerical solution. While minimizing
the $\gamma$-divergence leads to a soft rejection in the sense that the
estimate downweights the deviant points, the $q$-Gaussian {mixture}
helps set a rejection region when the deviation gets too large. Both of
these factors resist outliers and contribute robustness to our
clustering algorithm. To execute the self-updating procedure, we start
with treating each individual data point as the cluster representative
of a singleton cluster and, in each iteration, we update the~cluster
representatives through the derived estimating equations until all the
representatives converge. This self-updating procedure ensures that
neither knowledge of the number of clusters nor random initial centers
are required.

To investigate how $\gamma$-SUP would perform when applied to cryo-EM
images, we tested two sets of cryo-EM images. The first set, consisting
of noisy simulated RNA polymerase II images of different views
projected from a defined orientation, was chosen in order to decouple
the alignment issues from clustering issues, allowing for quantitative
comparison between $\gamma$-SUP and other clustering methods. The
second set consisted of 5000 real cryo-EM images of ribosome bound
with an elongation factor that locks it into a defined conformation.
For the test on the simulated data, both perfectly aligned cases and
misaligned cases were examined. $\gamma$-SUP did well in separating
different views in which the images were perfectly aligned and was able
to identify most of the deliberately misaligned images as outliers. For
the ribosome images, $\gamma$-SUP was successful in that the cluster
averages were consistent with the views projected from the known
ribosome structure.

The paper is organized as follows. Section~\ref{sec.review} reviews the
concepts of $\gamma$-divergence and the $q$-Gaussian distribution,
which are the core components of \mbox{$\gamma$-SUP}. In Section~\ref
{sec.method} we develop our $\gamma$-SUP clustering algorithm from the
perspective of the minimum $\gamma$-divergence estimation of the
$q$-Gaussian mixture model. The performance of $\gamma$-SUP is further
evaluated through simulations in Section~\ref{sec.examples}, and
through a set of real cryo-EM images in Section~\ref{sec.real}. The
paper ends with a conclusion in Section~\ref{sec.conclusion}.

\section{A review of \texorpdfstring{${\gamma}$}{gamma}-divergence and the ${q}$-Gaussian
distribution}
\label{sec.review}

In this section we briefly review the concepts of $\gamma$-divergence
and the $q$-Gaussian distribution, which are the key technical tools
for our $\gamma$-SUP clustering algorithm.

\subsection{\texorpdfstring{${\gamma}$}{gamma}-divergence}

The most widely used divergence of distributions is probably the
Kullback-Leibler divergence (KL-divergence) due to its connection to
maximum likelihood estimation (MLE). The $\gamma$-divergence, indexed
by a power parameter $\gamma> 0$, is a generalization of
KL-divergence. Let
\[
\M= \biggl\{f\dvtx0 < \int_\X f^{\gamma+ 1} < \infty,
f \ge0 \biggr\},
\]
where $f\dvtx{\cal X}\subset\mathbb{R}^n \mapsto\mathbb{R}^+$ is a
nonnegative function defined on ${\cal X}$. For simplicity, we assume
${\cal X}$ is either a discrete set or a connected region.
%
%
\begin{defn}[{[\citet{Fujisawa,Cichocki,Eguchi2}]}]
For $f,g\in\M$, define the $\gamma$-divergence $D_\gamma(\cdot\|
\cdot)$
and $\gamma$-cross entropy $C_\gamma(\cdot\|\cdot)$ as follows:
%
%
\begin{eqnarray}
\label{gammaDiv} D_\gamma(f\|g) = C_\gamma(f\|g) -
C_\gamma(f\|f)
\nonumber
\\[-8pt]
\\[-8pt]
\eqntext{\displaystyle\mbox{with } C_\gamma(f\|g) = -\frac{1}{\gamma(\gamma
+ 1)}
\int\frac{g^\gamma
(x)}{\|g\|_{\gamma+1}^\gamma} f(x) \,dx,}
\end{eqnarray}
where $\|g\|_{\gamma+1} = \{\int g^{\gamma+ 1}(x) \,dx\}^{1/(\gamma+
1)}$ is a normalizing constant.
\end{defn}

The $\gamma$-divergence can be understood as the divergence function
associated with a specific scoring function, namely, the
pseudospherical score [\citet{Good,Gneiting}]. The pseudospherical
score is given by $S(f, x) = f^{\gamma}(x) / \|f\|_{\gamma+
1}^\gamma$. The associated divergence function between $f$ and $g$
can be calculated from equation (7) in \citet{Gneiting} to be
\[
d(f\|g) = \int S(f, x) f(x) \,dx - \int S(g, x) f(x) \,dx = \gamma
(\gamma+ 1)
D_\gamma(f\|g).
\]
This implies that $d(\cdot\|\cdot)$ and $D_\gamma(\cdot\|\cdot)$ are
equivalent. Moreover, $D_\gamma(\cdot\|\cdot)$ can also be expressed as
a functional Bregman divergence [\citet{Frigyik}] by taking $\Phi
(f)=\|
f\|_{\gamma+ 1}$. The corresponding Bregman divergence is
\begin{eqnarray*}
D_\Phi(f\|g) &=& \Phi(f) - \Phi(g) - \delta\Phi[g, f-g] = \|f
\|_{\gamma+ 1} - \int\frac{g^\gamma(x)}{\|g\|_{\gamma+ 1}^\gamma
} f(x) \,dx
\\
&=& \gamma(\gamma+ 1) D_\gamma(f\|g),
\end{eqnarray*}
where $\delta\Phi[g, h]$ is the G\^ateaux derivative of $\Phi$ at $g$
along direction $h$. Note that $\|g\|_{\gamma+ 1}$ is a normalizing
constant so that the cross entropy enjoys the property of being
projective invariant, that is, $C_\gamma(f\|cg) = C_\gamma(f\|g)$,
$\forall c > 0$ [\citet{Eguchi2}]. By H\"older's inequality, it can
be shown that, for $f, g \in\Omega$ (defined below), $D_\gamma(f\|g)
\ge0$ and equality holds if and only if $g = \lambda f$ for some
$\lambda> 0$ [\citet{Eguchi2}]. Thus, by fixing a scale, for example,
\[
\Omega= \bigl\{f \in\M\dvtx\|f\|_{\gamma+ 1} = 1 \bigr\},
\]
$D_\gamma$ defines a legitimate divergence on $\Omega$. There are other
possible ways of fixing a scale, for example, $\Omega= \{f\in\M
\dvtx\int f(x) \,dx = 1 \}$.

In the limiting case, $\lim_{\gamma\to0} D_\gamma(f\|g) = D_0(f\|g) =
\int f(x) \ln\{f(x)/g(x)\} \,dx$, which gives the KL-divergence. The
MLE, which corresponds to the minimization of the KL-divergence
$D_0(\cdot\|\cdot)$, has been shown to be optimal in parameter
estimation in many settings in the sense of having minimum asymptotic
variance. This optimality comes with the cost that the MLE relies on
the correctness of model specification. Therefore, the MLE or the
minimization of the KL-divergence may not be robust against model
deviation and outlying data. On the other hand, the minimum $\gamma
$-divergence estimation is shown to be robust [\citet{Fujisawa}] against
data contamination. It is this robustness property that makes $\gamma
$-divergence suitable for the estimation of mixture components, where
each component is a local model [\citet{Mollah}].

\subsection{The ${q}$-Gaussian distribution}

The $q$-Gaussian distribution is a generalization of the Gaussian
distribution obtained by replacing the usual exponential function with
the $q$-exponential
\[
\exp_q(u) = \bigl\{1 + (1 - q)u \bigr\}_+^{{1}/{(1 - q)}}\qquad \mbox{where } \{x\}_+ = \max\{x, 0\}.
\]
In this article, we adopt $q < 1$, which corresponds to $q$-Gaussian
distributions with compact support. We will explain this necessary
condition of compact support later.

Let $\S_p$ denote the collection of all strictly positive definite
$p\times p$ symmetric matrices.
%
%
\begin{defn}[{[Modified from \citet{Amari,Eguchi2}]}]
\label{def.qGaussian}
For a fixed $q < 1 + \frac{2}{p}$, define the $p$-variate $q$-Gaussian
distribution $G_q(\mu,\Sigma)$ with parameters $\theta= (\mu,
\Sigma)
\in{\mathbb R}^p \times\S_p$ to have the probability density
function (p.d.f.),
%
%
\begin{equation}
\label{Gq.pdf} f_q(x; \theta) = \frac{c_{p,q}}{(\sqrt{2\pi})^p
\sqrt{|\Sigma|}}
\exp_q \bigl\{u(x; \theta) \bigr\}, \qquad x\in{\mathbb R}^p,
\end{equation}
where $u(x; \theta) = -\frac{1}{2} (x - \mu)^T \Sigma^{-1} (x - \mu)$
and $c_{p,q}$ is a constant so that $\int f_q(x;\break  \theta) \,dx = 1$.
\end{defn}

The constant $c_{p,q}$ is given below [cf. \citet{Eguchi2}]:
%
%
\begin{eqnarray}
c_{p,q} = \cases{ %
\displaystyle\frac{(1 - q)^{p/2}
\Gamma(1 + {p}/{2} + {1}/{(1-q)}
)} {
\Gamma(1 + {1}/{(1-q)} )}, &\quad
$\mbox{for } -\infty< q < 1,$
\vspace*{2pt}\cr
1, &\quad $\mbox{for } q \to1,$
\vspace*{2pt}\cr
\displaystyle\frac{(q - 1)^{p/2} \Gamma({1}/{(q - 1)} )} {
\Gamma({1}/{(q - 1)} - {p}/{2} )}, & \quad$\mbox{for } \displaystyle 1 < q < 1 + \frac
{2}{p}.$}
\end{eqnarray}
The class of the $q$-Gaussian distributions covers some well-known
distributions. In the limit as $q$ approaches 1, the $q$-Gaussian
distribution reduces to the Gaussian distribution. For $1 < q < 1 +
\frac{2}{p}$, the $q$-Gaussian distribution becomes the multivariate
$t$-distribution. This can be seen by setting $v = 2/(q-1)-p > 0$.
Then, $f_q(x; \theta)$ in \eqref{Gq.pdf} is proportional to
%
%
\begin{equation}
\label{t.pdf} \biggl\{1 + \frac{1}{v} (x - \mu)^T \biggl(
\frac{p + v}{v} \Sigma\biggr)^{-1} (x - \mu) \biggr
\}^{-{(p + v)}/{2}},
\end{equation}
which is exactly the p.d.f. of a $p$-variate $t$-distribution (up to a
constant term) with location and scale parameters $(\mu, \frac{p +
v}{v} \Sigma)$ and degrees of freedom $v$. Depending on the choice
of $q$, the support of $G_q(\mu, \Sigma)$ also differs. For $1 +
\frac
{2}{p} > q \ge1$ (i.e., for the Gaussian distribution and
$t$-distribution), the support of $G_q(\mu,\Sigma)$ is the entire~$\mathbb{R}^p$. For $q < 1$, however, the support of $G_q(\mu,\Sigma)$
is compact and depends on $q$ in the form
%
%
\begin{equation}
\label{domain.q} \biggl\{x\dvtx(x - \mu)^T \Sigma^{-1} (x -
\mu) < \frac{2}{1 - q} \biggr\}.
\end{equation}
Thus, choosing $q < 1$ leads to zero mutual influence between clusters
in our clustering algorithm. Note that if $X \sim G_q(\mu, \Sigma)$
with $q < 1 + \frac{2}{p+2}$,\footnote{For $ q < 1 + \frac{2}{p +
\nu
}$, it ensures the existence of the $\nu$th moment of $X$.} then
$E(X) = \mu$ and $\cov(X) = \frac{2}{2 + (p + 2)(1 - q)} \Sigma$.

\subsection{Minimum \texorpdfstring{${\gamma}$}{gamma}-divergence for estimating a ${q}$-Gaussian}
\label{sec.gamma_div_q_Gaussian}

The $\gamma$-divergence is a discrepancy measure for two functions in
$\M$. Its minimum can then be used as a criterion to approximate an
underlying p.d.f. $f$ from a certain model class $\M_\Theta$ parameterized
by $\theta\in\Theta\subset{\mathbb R}^m$. It has been deduced that
minimizing $D_\gamma(f\|g)$ over $g$ is equivalent to minimizing the
$\gamma$-loss function [\citet{Fujisawa}]
%
%
\begin{equation}
\label{loss} L_{\gamma, f}(g) = -\frac{1}{\gamma} \ln\biggl\{\int
g^{\gamma}(x) f(x) \,dx \biggr\} + \frac{1}{\gamma+ 1} \ln\biggl\{
\int
g^{\gamma+ 1}(x) \,dx \biggr\}.
\end{equation}
Thus, at the population level, $f$ is estimated by
%
%
\begin{equation}
\label{theta_star} f^*= \argmin_{g \in\M_\Theta} D_\gamma(f\|g) =
\argmin_{g \in\M_\Theta
} L_{\gamma,f}(g).
\end{equation}
At this moment, we consider $\M_\Theta$ to be the family of
$q$-Gaussian distributions $G_q(\mu, \Sigma)$ with $\theta= (\mu,
\Sigma)$. Then, for any given values of $\gamma$ and $q$, the loss
function $L_{\gamma, f}(g)$ evaluated at $g(x) = f_q(x; \theta) \in
\M
_\Theta$ becomes
\begin{eqnarray*}
L_{\gamma, f} \bigl\{f_q(x; \theta) \bigr\} 
&=& -\frac{1}{\gamma} \ln\biggl[\int f(x)
\biggl\{\frac{\exp_q(u(x; \theta))^{\gamma+ 1}} {
\int\{\exp_q (u(v; \theta)) \}^{\gamma+ 1} \,dv} \biggr\}
^{{\gamma}/{(\gamma+ 1)}} \,dx \biggr]
\\
&=& -\frac{1}{\gamma} \ln\biggl[\int f(x)
\biggl\{f_{{(\gamma+ q)}/{(\gamma+ 1)}}\biggl(x;
\mu,\frac{1}{\gamma+1} \Sigma\biggr) \biggr\}^{{\gamma}/{(\gamma
+ 1)}} \,dx \biggr].
\end{eqnarray*}
Direct calculation then gives that minimizing $L_{\gamma, f}\{f_q(x;
\theta)\}$ over possible values of $\theta$ is equivalent to maximizing
%
%
\begin{equation}
\label{loss.prop} \int f(x) |\Sigma|^{-({1}/{2})({\gamma}/{(\gamma+
1)})} \bigl[\exp_q\bigl
\{u(x; \theta)\bigr\} \bigr]^{\gamma} \,dx.
\end{equation}
For high-dimensional data, however, it is impractical to estimate the
covariance matrix $\Sigma$ and its inverse. Since our main interest is
to find cluster centers, we employ $\Sigma= \sigma^2 I_p$ as our
working model. By taking the derivative of \eqref{loss.prop} with
respect to $\mu$, we get the stationary equation for the maximizer
$\mu
^*$ for any fixed $\sigma^2$:
%
%
\begin{eqnarray}
\label{mu} \mu^* &=& \frac{\int x f(x) [\exp_{q} \{u(x; \mu^*,
\sigma^2)\}
]^{\gamma- (1 - q)} \,dx} {
\int f(x) [\exp_{q} \{u(x; \mu^*, \sigma^2)\} ]^{\gamma-
(1 - q)} \,dx}
\nonumber
\\[-8pt]
\\[-8pt]
\nonumber
&=& \frac{\int x w(x; \mu^*, \sigma^2) \,dF(x)} {\int
w(x; \mu^*, \sigma
^2) \,dF(x)},
\end{eqnarray}
where
$w(x; \mu^*, \sigma^2) = [\exp_{q}\{u(x; \mu^*, \sigma^2)\}
]^{\gamma- (1 - q)}$\vspace*{1pt} is the weight function and $F(x)$ is the
cumulative distribution function corresponding to $f(x)$.

Given the observed data $\{x_i\}_{i=1}^n$, the sample analogue of
$\mu^*$ can be obtained naturally by replacing $F(x)$ in \eqref{mu}
with the empirical distribution function $\hat F(x)$ of
$\{x_i\}_{i=1}^n$. This gives the stationarity condition for $\mu^*$
at the
sample level:
%
%
\begin{equation}
\label{mu.data} \mu^* = \frac{\sum_{i=1}^n x_i w(x_i; \mu^*,
\sigma^2)}{\sum_{i=1}^n
w(x_i; \mu^*, \sigma^2)}.
\end{equation}
One can see that the weight function $w$ assigns the contribution of
$x_i$ to $\mu^*$. Thus, a robust estimator should encourage the
property that smaller weight is given to those $x_i$ farther away from
$\mu^*$ and zero weight to extreme outliers. These can be achieved by
choosing proper values of $(\gamma, q)$ in $w$. In particular, when $q
< 1$, we have from \eqref{domain.q} that
%
%
\begin{equation}
\label{hard_range} w\bigl(x; \mu^*, \sigma^2\bigr) = \cases{ %
\displaystyle\biggl(1 - \frac{1 - q}{2\sigma^2}\bigl \|x - \mu^*\bigr\|_2^2
\biggr)^{ {(\gamma- (1 - q))}/ {(1-q)}}, \vspace*{2pt}\cr
\hspace*{32pt}\mbox{for }\displaystyle
\bigl\|x - \mu^*
\bigr\|_2^2 < \frac{2\sigma
^2}{1 - q},
\vspace*{2pt}\cr
0, \qquad\mbox{for } \displaystyle\bigl\|x - \mu^*\bigr\|_2^2 \ge
\frac{2\sigma^2}{1 - q}.}
\end{equation}
That is, data points outside the range $\mu^*\pm\frac{2\sigma^2}{1
- q}$ do not have any influence on~$\mu^*$. Note also that when
$\gamma= 1 - q$, then $w(x; \mu^*, \sigma^2) = 1$ and, thus, $\mu^*$
in \eqref{mu.data} becomes the sample mean $n^{-1} \sum_{i=1}^n
x_i$, which is not robust to outliers. This fact suggests that we
should use a
$\gamma$ value that is greater than $1 - q$.

\section{\texorpdfstring{$\gamma$}{gamma}-SUP}
\label{sec.method}

In this section we introduce our clustering method, $\gamma$-SUP, which
is derived from minimizing $\gamma$-divergence under a $q$-Gaussian
mixture model.

\subsection{Model specification and estimation}
\label{sec.gammaSUP.method}

Suppose we have collected data $\{x_i\}_{i=1}^n$ with empirical
probability mass $\hat f(x)$ and empirical c.d.f. $\hat F(x) = \frac{1}{n}
\sum_{j=1}^n I(x_j \le x)$, where ``$\le$'' is understood
componentwise. The goal is to group them into $K$ clusters, where $K$
is unknown and should be determined from the data. Assume that $f$ is a
mixture of $K$ components with p.d.f.
%
%
\begin{equation}
\label{mixture_f} f(x) = \sum_{k=1}^K
\pi_k f_q(x; \theta_k),
\end{equation}
where each component is modeled by a $q$-Gaussian distribution
indexed by $\theta_k = (\mu_k,\sigma^2)$. Most model-based
clustering approaches (e.g., those assuming a Gaussian mixture
model) aim to learn the whole model $f$ by minimizing a divergence
(e.g., KL-divergence) between $f$ and the empirical probability mass
$\hat f$. They therefore suffer the problem of having to specify the
number of components $K$ before implementation. To overcome this
difficulty, instead of minimizing $D_\gamma(\hat f \| f)$ to learn
$f$ directly, we consider learning each component $f_q(\cdot;
\theta_k)$ of $f$ separately through the minimization problem
%
%
\begin{equation}
\label{gammasup_estimate} \min_\theta D_\gamma\bigl\{\hat f \|
f_q(\cdot; \theta)\bigr\}.
\end{equation}
The validity of \eqref{gammasup_estimate} to learning all
components of $f$ relies on the locality of $\gamma$-divergence, as
shown in Lemma 3.1 of \citet{Fujisawa}. The authors have proven that,
at the population level, $D_\gamma\{f \| f_q(\cdot; \theta)\}$ is
approximately proportional to $D_\gamma\{f_q(\cdot; \theta_k) \|
f_q(\cdot; \theta)\}$, provided that the model $f_q(x; \theta)$ and
the remaining components $\{f_q(x; \theta_\ell) \dvtx\ell\ne k\}$ are
well separated. We also refer the readers to \citet{Mollah} for a
comprehensive discussion about the locality of $\gamma$-divergence.
Consequently, we are motivated to find all local minimizers of
\eqref{gammasup_estimate}, each of which corresponds to an estimate
of one component of $f$. Moreover, the number of local minimizers
provides an estimate of $K$. A detailed implementation algorithm
that finds all local minimizers and estimates $K$ is introduced in
the next subsection.

\subsection{Implementation: Algorithm and tuning parameters}

We have shown in Section~\ref{sec.gamma_div_q_Gaussian} that, for
any given $\sigma^2$, solving \eqref{gammasup_estimate} is
equivalent to finding the cluster center $\mu^*$ that satisfies the
stationary equation \eqref{mu.data}. Starting with a set of initial
cluster centers $\{\hat\mu^{(0)}_i\}$ indexed by $i$, we consider
the following fixed-point algorithm to solve \eqref{mu.data}:
%
%
\begin{equation}
\label{sup0.nb} \hat\mu^{(\ell+ 1)}_i = \frac{\int x w(x; \hat\mu
^{(\ell)}_i, \sigma^2) \,d\hat F(x)} {
\int w(x; \hat\mu^{(\ell)}_i, \sigma^2) \,d\hat F(x)}, \qquad\ell=
0, 1, 2, \ldots.
\end{equation}
Multiple initial centers are necessary to find multiple solutions of
\eqref{mu.data}. To avoid the problem of random initial centers, in
this paper we consider the natural choice
%
%
\begin{equation}
\label{initial} \bigl\{\hat\mu^{(0)}_i = x_i
\bigr\}_{i=1}^n.
\end{equation}
Other choices are possible, but \eqref{initial} gives a straightforward
updating path $\{\hat\mu_i^{(\ell)} \dvtx\ell= 0, 1, 2, \ldots\}$ for
each observation $x_i$. At convergence, the distinct values of $\{\hat
\mu^{(\infty)}_i\}_{i=1}^n$ provide estimates of the cluster centers
$\{
\mu_k\}_{k=1}^K$ and the number of clusters. Moreover, cases whose
updating paths converge to the same cluster center are clustered
together. Though derived from a minimum $\gamma$-divergence
perspective, we note that \eqref{sup0.nb} combined with \eqref{initial}
has the same form as the mean-shift clustering [\citet{Fukunaga}] for
mode seeking. It turns out that clustering through \eqref
{sup0.nb}--\eqref{initial} shares the same properties as mean-shift
clustering. Our $\gamma$-SUP framework provides the mean-shift
algorithm an information theoretic justification, namely, it minimizes
the $\gamma$-divergence under a $q$-Gaussian mixture model. We call
\eqref{sup0.nb}--\eqref{initial} the ``nonblurring $\gamma
$-estimator,'' to distinguish it from our main proposal introduced in
the next paragraph.

While the nonblurring mean-shift updates the cluster centers with
the original data points being fixed [which corresponds to a fixed
$\hat F$ in \eqref{sup0.nb} over iterations], the blurring
mean-shift [\citet{ChengY}] is a variant of the nonblurring
mean-shift algorithm, which updates the cluster centers and moves
(i.e., blurs) the data points simultaneously. \citet{Shiu} proposed
self-updating process (SUP) as a clustering algorithm. The blurring
mean-shift can be viewed as an SUP with a homogeneous updating rule
(SUP allows nonhomogeneous updating). It has been reported
[\citet{Shiu}] that SUP possesses many advantages, especially in the
presence of outliers. Thus, we are motivated to implement the
minimum $\gamma$-divergence estimation via an SUP-like algorithm and
call it $\gamma$-SUP. In particular, the $\gamma$-SUP algorithm is
constructed by replacing $\hat F(x)$ in \eqref{sup0.nb} with $\hat
F^{(\ell)}(x) = \frac{1}{n} \sum_{j=1}^n I(\hat\mu_j^{(\ell)} \le
x)$, to reflect its nature in updating blurred data as model
representatives over iterations. That is, after one iteration, we
believe that the blurred data is more representative of the
population than the previous one and, thus, the blurred data is
treated as the new sample to be entered into the next iteration.
This gives the self-updating process of $\gamma$-SUP:
%
%
\begin{equation}
\label{sup0} \hat\mu^{(\ell+1)}_i = \frac{\int x w(x; \hat\mu
^{(\ell)}_i, \sigma^2) \,d\hat F^{(\ell)}(x)} {
\int w(x; \hat\mu^{(\ell)}_i, \sigma^2) \,d\hat F^{(\ell)}(x)},\qquad
\ell=
0, 1, 2, \ldots.
\end{equation}
The update \eqref{sup0} can be expressed as, for $i = 1, \ldots, n$,
%
%
\begin{equation}
\label{sup1} \hat\mu^{(1)}_i = \frac{\sum_{j=1}^n w_{ij}^{(0)}
\hat\mu^{(0)}_j} {
\sum_{j=1}^n w_{ij}^{(0)}}
\rightarrow\hat\mu^{(2)}_i = \frac{\sum_{j=1}^n w_{ij}^{(1)} \hat
\mu^{(1)}_j } {
\sum_{j=1}^n w_{ij}^{(1)}} \rightarrow
\cdots\rightarrow\hat\mu^{(\infty)}_i,
\end{equation}
where
%
%
\begin{eqnarray}
\label{weight} w_{ij}^{(\ell)}&=& \biggl\{\exp_q
\biggl(-\frac{1}{2\sigma^2} \bigl\llVert\hat\mu^{(\ell)}_i -
\hat\mu^{(\ell)}_j \bigr\rrVert_2^2
\biggr) \biggr\}^{\gamma-(1-q)}
\nonumber
\\[-8pt]
\\[-8pt]
\nonumber
& =& \exp_{1-s} \biggl(-\frac
{1}{\tau^2}
\bigl\llVert\hat\mu^{(\ell)}_i - \hat\mu^{(\ell)}_j
\bigr\rrVert_2^2 \biggr)
\end{eqnarray}
with the scale parameter $\tau$ and the model parameter $s$ being
%
%
\begin{equation}
\label{tau_s} \tau= \sigma\sqrt{\frac{2}{\gamma- (1 - q)}} \quad\mbox{and}\quad s =
\frac{1 - q}{\gamma- (1 - q)} > 0.
\end{equation}
Here $s > 0$ is a consequence of choosing $q<1$ and $\gamma> 1-q$ as
mentioned in the end of Section~\ref{sec.gamma_div_q_Gaussian}. Thus,
$\gamma$-SUP involves only $(s, \tau)$ as the tuning parameters. It has
been found in our numerical studies that $\gamma$-SUP is quite
insensitive to the choice of $s$ and that $\tau$ plays the decisive
role in the performance of $\gamma$-SUP. We thus suggest a choice of a
small positive value of $s$, say, 0.025, in practical implementation. A
phase transition plot (Figure~\ref{ClassNumberScale}) will be
introduced to determine $\tau$ in Section~\ref{sec.examples}.

It can be seen from \eqref{sup1} that, in updating
$\hat\mu^{(\ell)}_i$ in the $\ell$th iteration, $\gamma$-SUP takes
a weighted average over the candidate model representatives
$\{\hat\mu_j^{(\ell)}\}_{j=1}^n$ according to the weights
$\{w_{ij}^{(\ell)}\}_{j=1}^n$. Due to the weights in (\ref{weight})
being nonnegative
and decreasing with respect to the distance $\|\hat\mu^{(\ell)}_i -
\hat\mu^{(\ell)}_j\|_2$ and the compact support of the $q$-Gaussian
distribution, the
convergence of $\gamma$-SUP is assured [\citet{Chen2}]. We can express
the weights \eqref{weight} as
%
%
\begin{equation}
\label{weight3} w_{ij}^{(\ell)} = \exp_{1-s} \bigl(-
\bigl\llVert\tilde\mu^{(\ell)}_i -\tilde\mu^{(\ell)}_j
\bigr\rrVert_2^2 \bigr) \qquad\mbox{with } \tilde
\mu_i^{(\ell)} = \hat\mu_i^{(\ell)} / \tau.
\end{equation}
As a result, $\gamma$-SUP starts with $n$ (scaled) cluster centers
$\{\tilde\mu^{(0)}_i = x_i / \tau\}_{i=1}^n$, which avoids the
problem of random initial centers. Eventually, $\gamma$-SUP
converges to certain $K$ clusters, where $K$ depends on the {tuning
parameters $(s, \tau)$}, but otherwise is data-driven. Moreover, we
have the cluster representatives $\{\hat\mu_i^{(\infty)}\}_{i=1}^n$,
which contain $K$ distinct points denoted by $\{\hat\mu_k = \tau
\tilde\mu_k\}_{k=1}^K$. The corresponding cluster membership for
each data point is denoted by $\{c_i\}_{i=1}^n$. The detailed
algorithm of $\gamma$-SUP \eqref{initial}--\eqref{sup0} is
summarized in Table~\ref{SUP}. Note that in our proposal, we ignore
the estimation of $\sigma^2$. The main reason is that $\sigma^2$ is
absorbed into the scale parameter $\tau$ defined in \eqref{tau_s},
and a phase transition plot can be used to select $\tau$ directly.

%
\begin{table}
\caption{$\gamma$-SUP clustering algorithm}
\label{SUP}
\begin{tabular*}{\textwidth}{@{\extracolsep{\fill}}l@{}}
\hline
\textbf{Inputs:}\quad Data matrix $X\in\mathbb{R}^{n\times p}$, $n$
instances with $p$ variables;\\
\hphantom{\textbf{Inputs:}\quad }Tuning parameters $(s, \tau)$. \\
\textbf{Outputs:} Number of clusters $K$ and cluster centers
$\{\hat\mu_k\}_{k=1}^K$; \\
\hphantom{\textbf{Outputs:} }Cluster membership assignment $\{c_i\}
_{i=1}^n$ for each of $\{x_i\}_{i=1}^n$. \\
\hline
\textbf{begin} \\
\hspace{0.8cm} $\Iter\leftarrow0$ \\
\hspace{0.8cm} \textbf{start with:} $\tilde\mu_i \leftarrow x_i /
\tau$,
$i=1, \ldots, n$ \\
\hspace{0.8cm} \textbf{repeat}\\
\hspace{1.6cm} \textbf{for $i = 1 \dvtx n$} \\
\hspace{2.0cm} $w_{ij} \leftarrow\exp_{1 - s} (-\llVert\tilde\mu_i
- \tilde\mu_j \rrVert_2^2 )$, $j = 1,\ldots, n$ \\
\hspace{2.0cm} $z_{i} \leftarrow\sum_{j=1}^{n} \frac{w_{ij}}{\sum
_{k=1}^{n} w_{ik}} \tilde\mu_{j}$ \\
\hspace{1.6cm} \textbf{end} \\
\hspace{1.6cm} $\tilde\mu_{i} \leftarrow z_{i}$, $i = 1, \ldots, n$
\\
\hspace{1.6cm} $\Iter\leftarrow\Iter+ 1$ \\
\hspace{0.8cm} \textbf{until} convergence\\
\hspace{0.8cm} \textbf{output} distinct cluster centers $\{\tau
\tilde
\mu_i, 1\le i\le n$\} and cluster membership \\
\textbf{end} \\
\hline
\end{tabular*}
\tabnotetext[]{}{\textit{Note}:
The parameter $\tau$ is linearly proportional to the
influence region radius that defines the similarity inside a
cluster.}
\end{table}

%

%
\begin{figure}

\includegraphics{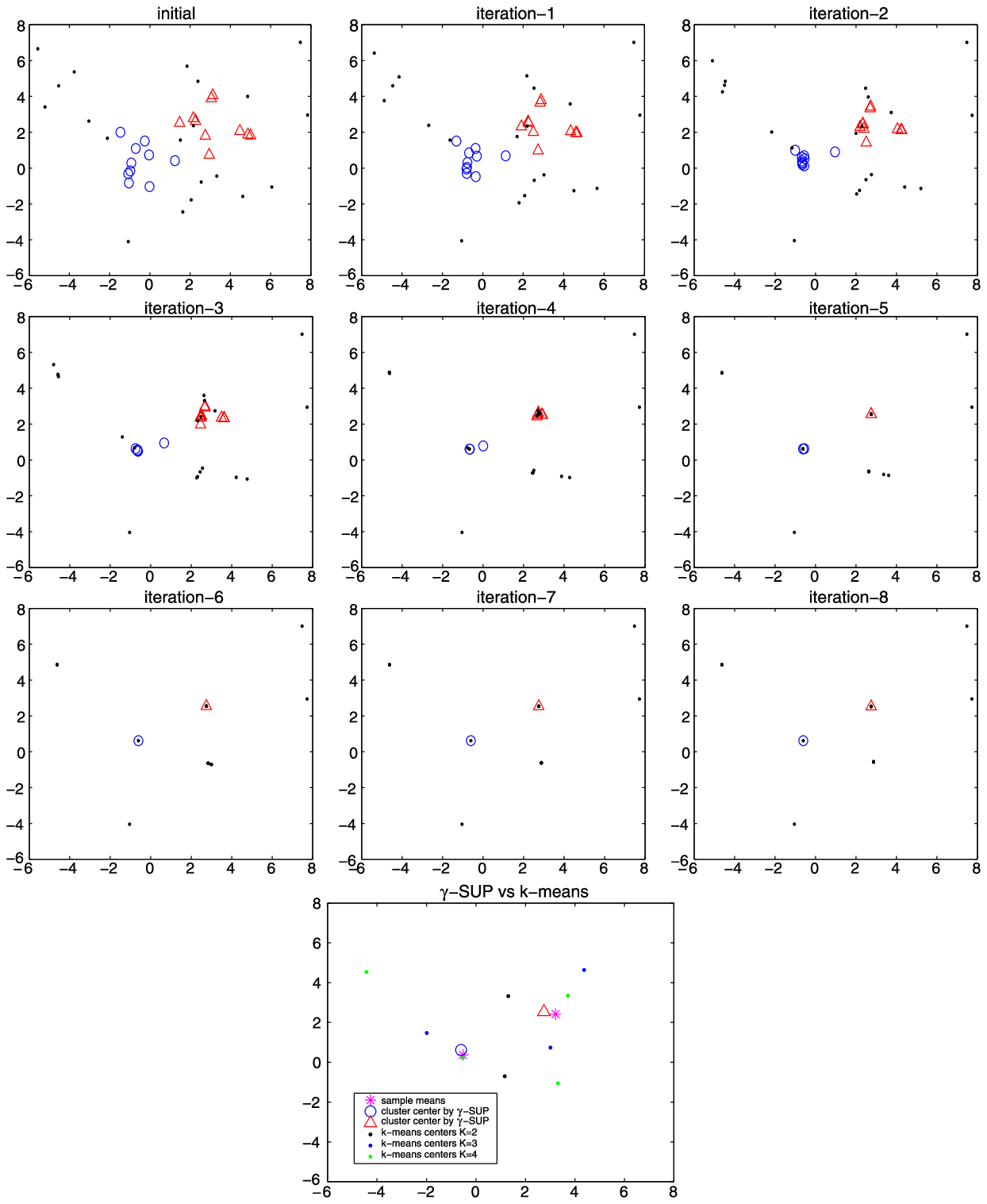}

\caption{How data points move by $\gamma$-SUP. The process stops at
the $8${th} iteration. Bottom row:
comparison with the sample means of the true clusters and $k$-means
centers under $K=2,3,4$.}\label{fig.exp1}
\end{figure}

A toy example to illustrate how data points move by $\gamma$-SUP is
presented in Figure~\ref{fig.exp1}. Two clusters with 10 data points
each are sampled from the standard normal distributions centered at
$(0,0)$ and $(2.355, 2.355)$, respectively. Another 20 isolated
noise points are added surrounding these two clusters. The first
plot (upper left) shows the initial positions of these 40 points.
Then each data point is updated (blurred) according to the weighted
average of its neighboring points. After the $8$th iteration, no
(blurred) data points move any more and the algorithm stops. Data
points with the same final position are assigned to the same
cluster. There are seven clusters at the end of the self-updating process.
Two cluster centers are close to the true means of the normal mixture.
The other five cluster centers are formed by noise data.
Data points sampled from the normal mixture are correctly merged into
their target
clusters. Some noise points close to these two clusters move into
them, while other noise points are merged into five clusters. At the
bottom row of Figure~\ref{fig.exp1}, we have also provided a plot for
comparison with sample means and $k$-means centers. The sample means
were computed with the information of the true cluster labels and
the $k$-means centers were computed under given $K=2,3,4$,
respectively. One can see that, in the presence of outliers, cluster
centers from $\gamma$-SUP can still be close to the sample means,
while this is not the case for $k$-means under every $K$.

\subsection{Characteristics of \texorpdfstring{$\gamma$}{gamma}-SUP}
\label{sec.gammaSUP_more}

Similar to \eqref{sup1}, the nonblurring $\gamma$-\break estimator~\eqref
{sup0.nb}--\eqref{initial} can be expressed as
%
%
\begin{equation}
\label{sup1.nb} \hat\mu^{(1)}_i = \frac{\sum_{j=1}^n w_{ij}^{*(0)}
x_j}{\sum_{j=1}^n
w_{ij}^{*(0)}}
\rightarrow\hat\mu^{(2)}_i = \frac{\sum_{j=1}^n w_{ij}^{*(1)}
x_j}{\sum_{j=1}^n
w_{ij}^{(*1)}} \rightarrow
\cdots\rightarrow\hat\mu^{(\infty)}_i,
\end{equation}
where $w_{ij}^{*(\ell)} = \exp_{1 - s} (-\frac{1}{\tau^2} \llVert
x_j - \hat\mu^{(\ell)}_i \rrVert_2^2 )$. Comparing \eqref{sup1}
and \eqref{sup1.nb}, the constructed cluster centers from $\gamma$-SUP
and the nonblurring $\gamma$-estimator are both weighted averages with
the weights $w_{ij}^{(\ell)}$ and $w_{ij}^{*(\ell)}$, respectively. The
principle of downweighting is important for robust model fitting
[\citet{Basu,Field,Windham}]. We emphasize that the downweighting by
$w_{ij}^{(\ell)}$ in \eqref{sup1} is \emph{with respect to models},
since each cluster center $\hat\mu_i^{(\ell)}$ is a weighted average of
$\{\hat\mu_i^{(\ell)}\}_{i=1}^n$. On the other hand, downweighting by
$w_{ij}^{*(\ell)}$ in \eqref{sup1.nb} is \emph{with respect to data},
since each cluster center $\hat\mu_i^{(\ell)}$ is a weighted average of
the original data $\{x_i\}_{i=1}^n$. The same concept can also be
observed in the difference between the nonblurring mean-shift and the
blurring mean-shift. It has been shown that the blurring mean-shift has
a faster rate of convergence [\citet{Carreira,Chen2}]. To our
knowledge, however, there is very little statistical evaluation for
these two downweighting schemes in the literature. As will be
demonstrated in Section~\ref{sec.examples}, from a statistical
perspective, downweighting with respect to models is more efficient
than with respect to data in estimating the mixture model, which
further supports the usage of $\gamma$-SUP in practice.

\section{Numerical study}
\label{sec.examples}

\subsection{Synthetic data}

\begin{figure}

\includegraphics{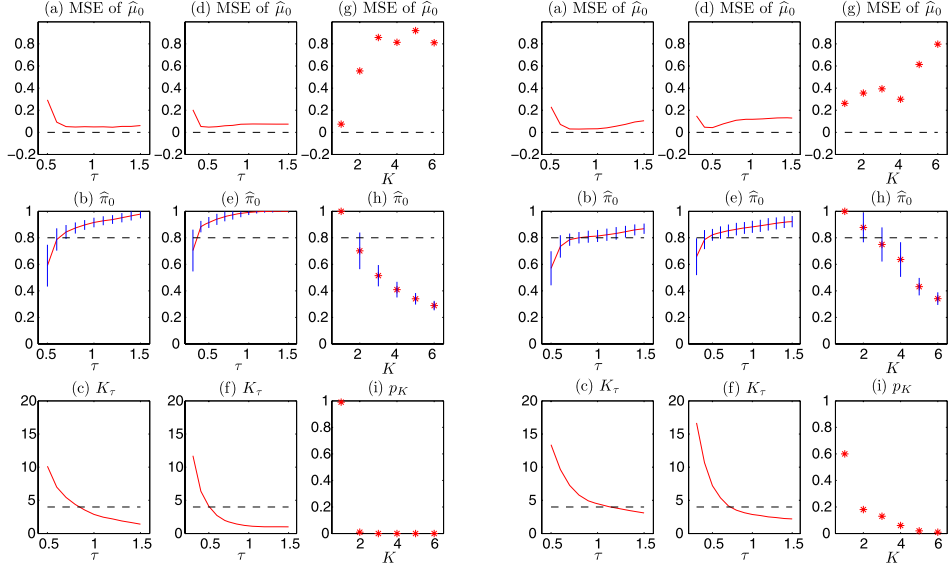}

\caption{Simulation results for the mixture model in \protect\eqref
{sim2.R1} with
$(c, \pi_0) = (2, 0.8)$ (the left panel) and $(c, \pi_0) = (4, 0.8)$
(the right panel). Results are
reported for $\gamma$-SUP
(the first column) and the nonblurring $\gamma$-estimator (the second
column) at $s=0.025$ and various $\tau$ values,
and $k$-means (the third column) at various $K$ values. \textup{(a)},
\textup{(d)}, \textup{(g)}:
MSE of $\hat\mu_0$;
\textup{(b)}, \textup{(e)}, \textup{(h)}: means of $\hat\pi_0$,
where the
vertical bars represent standard deviations; \textup{(c)}, \textup{(f)}:
means of $K_\tau$;
\textup{(i)}: means of $p_K$. The horizontal dashed line represents the
target value.}
\label{fig.sim2.R1}
\end{figure}

We show by simulation that $\gamma$-SUP is more efficient in model
parameter estimation than both the nonblurring $\gamma$-estimator
and the $k$-means estimator. Random samples of size 100 are
generated from a 4-component normal mixture model with p.d.f.
%
%
\begin{equation}
\label{sim2.R1} \pi_0 f(x; \mu_0, \Sigma) + \sum
_{k=1}^3 \frac{1-\pi_0}{3} f(x;
\mu_k, \Sigma),
\end{equation}
where $\Sigma= I_2$, $\mu_0 = (0,0)^T$, $\mu_1 = (c,c)^T$, $\mu_2 =
(c, -2c)^T$, and $\mu_3 = (-c, 0)^T$ for some $c$. We set $\pi_0 =
0.8$ and treat $f(x; \mu_0, \Sigma)$ as the relevant component of
interest, while $\{f(x; \mu_k, \Sigma) \dvtx k = 1, 2, 3\}$ are noise
components. We apply $\gamma$-SUP and the nonblurring
$\gamma$-estimator (both with $s = 0.025$) and $k$-means to
estimate $(\mu_0, \pi_0)$ by using the largest cluster center and
cluster size fraction as $\hat\mu_0$ and $\hat\pi_0$, respectively.
The simulation results with 100 replicates under different $c$
values are placed in Figure~\ref{fig.sim2.R1}, which includes the
MSE of $\hat\mu_0$ and the mean of $\hat\pi_0$ plus/minus one
standard deviation. To evaluate the sensitivity of each method to
the selection of $\tau$ or $K$, we report $K_\tau$ (the average
selected number of components under $\tau$) for $\gamma$-SUP and the
nonblurring $\gamma$-estimator, and report $p_K$, the
probability of selecting $K$ components by the gap-statistic
[\citet{Tibshirani}], for $k$-means.

For the case of closely spaced components, where $c = 2$, although
$\gamma$-SUP and the nonblurring $\gamma$-estimator perform
similarly, $\gamma$-SUP produces smaller MSE in a wider range of
$\tau$ values. Moreover, the mean of $\hat\pi_0$ from $\gamma$-SUP
is closer to the target value $0.8$. On the other hand, $k$-means
provides unsatisfactory results for $K>1$, indicating that $k$-means
is sensitive to the selection of $K$ when components are not well
separated. The superiority of $\gamma$-SUP again has been found in the
case of a moderate distance, $c = 4$, among the components. In this
situation, $\gamma$-SUP still provides good estimates of $(\mu_0,
\pi_0)$ over a wide range of $\tau\in[0.6, 1]$, while the
selection of $\tau$ becomes more critical for the nonblurring
$\gamma$-estimator. The minimum MSE of the nonblurring
$\gamma$-estimator happens only around $\tau= 0.5$. Since $\tau$
should be determined from the data, this fact supports the
applicability of $\gamma$-SUP in practical implementation. The
$k$-means algorithm
can produce satisfactory results at the true number of components $K
= 4$, although its MSE is still the largest among the three methods.
However, the gap-statistic does not support the choice of $K = 4$,
which increases the difficulty of using a $k$-means
with gap-statistic approach for clustering unbalanced-sized components.

%
\begin{figure}

\includegraphics{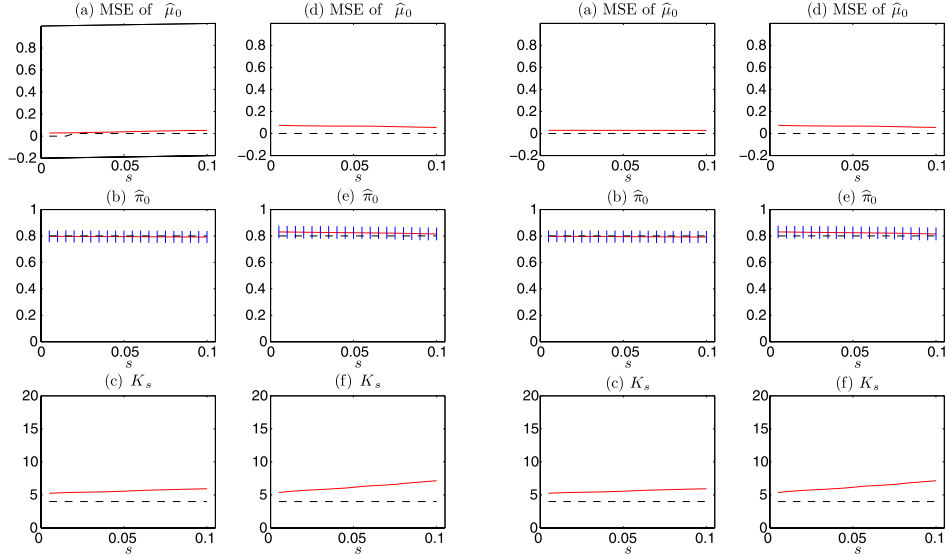}

\caption{Simulation results for the mixture model in \protect\eqref
{sim2.R1} with
$(c, \pi_0) = (2, 0.8)$ (the left panel) and $(c, \pi_0) = (4, 0.8)$
(the right panel). Results are
reported for $\gamma$-SUP (the first column) and the nonblurring
$\gamma$-estimator
(the second column) at $\tau= 0.6$ and $s \in[0.005, 0.1]$. \textup{(a)},
\textup{(d)}: MSE of $\hat\mu_0$; \textup{(b)}, \textup{(e)}:
means of $\hat\pi_0$, where the
vertical bars represent standard deviations; \textup{(c)}, \textup
{(f)}: mean selected
number of components
$K_s$ at $s$. The horizontal dashed line represents the target value.}
\label{fig.sim2_s.R1}
\end{figure}

We now evaluate the influence of $(s, \tau)$ to $\gamma$-SUP. The
same simulation studies with $\tau= 0.6$ and $s \in[0.005, 1]$
are conducted. One can see from Figure~\ref{fig.sim2_s.R1} that the
performance of $\gamma$-SUP and the nonblurring $\gamma$-estimator is
quite insensitive to the value of
$s$ and that $\gamma$-SUP is even a bit better, which confirms that
the critical tuning parameter is the scale parameter $\tau$. As to
the selection of $\tau$, one can find an elbow-pattern (or phase
transition) of $K_\tau$ in Figures~\ref{fig.sim2.R1}(c) and (f).
This phenomenon suggests the necessity of choosing a $\tau$ value of
which the number of constructed components becomes stable. It can be
seen from Figure~\ref{fig.sim2.R1}(a)--(b) and (d)--(e) that the
suggested rule corresponds to reasonable performance for
$\gamma$-SUP and the nonblurring $\gamma$-estimator, where the best
result is still from $\gamma$-SUP. We will further examine this
selection criterion in the next study using simulated cryo-EM
images. Moreover, as discussed in Section~\ref{sec.gammaSUP_more},
$\gamma$-SUP has a faster rate of convergence. Therefore, the
computational time of $\gamma$-SUP is shorter than the nonblurring
$\gamma$-estimator. In summary, $\gamma$-SUP performs better under
different settings and is less sensitive to $s$-selection.

\subsection{Simulated cryo-EM images}
\label{cryo_em}

We use simulated cryo-EM images to demonstrate the superiority of
$\gamma$-SUP in clustering homogeneous images while isolating
misaligned images. A total of 128 distinct 2D images with $100 \times
100$ pixels were generated by projecting the X-ray crystal structure of
RNA polymerase II filtered to 20 Angstroms in equally spaced
(angle-wise) orientations.\footnote{Data source: The X-ray model of RNA
polymerase II is from the Protein Data Bank (PDB: 1WCM).} Each image
was then convoluted with the electron microscopy contrast transfer
function (defocus 2 $\mu$m). Finally, 6400 images were randomly
sampled with replacement from these 128 projections with i.i.d.
Gaussian noise $N(0, \sigma_\varepsilon^2)$ added, where $\sigma
_\varepsilon
= 40, 50, 60$, so that the corresponding SNRs are $0.19, 0.12, 0.08$,
which reflect the low SNR nature of cryo-EM images. This procedure of
simulating cryo-EM images is commonly used in the cryo-EM community
[\citet{Chang,Singer,Sorzano,Hall}]. To reflect the nature of the
experiment, we consider two scenarios:
\begin{longlist}[1.]
\item[1.]
All the images are perfectly aligned.
\item[2.]
A portion of the images are misaligned.
\end{longlist}
The misaligned images are treated as outliers and should be identified
individually without clustering with other correctly aligned images.
Before entering the clustering algorithm, we conducted dimension
reduction on this image data set. Instead of using principal component
analysis (PCA), we applied multilinear principal component analysis
[MPCA, \citet{Lu,Hung}] to reduce the dimension from $100 \times100$
to $10 \times10$, as MPCA has been shown to be more efficient in
dimension reduction for array data such as image sets [\citet{Hung}].
The extracted low-dimensional MPCA factor loadings then enter the
clustering analysis.

We have tested $\gamma$-SUP, $k$-means, clustering 2D (CL2D)
[\citet{Sorzano}] and $k$-means$^+$, a variant of the latter two, for
comparison. Given a prespecified number of clusters $K$, unlike
$k$-means which separates the data into $K$ clusters directly, CL2D
bisects the clusters iteratively until $K$ clusters are constructed.
During the clustering process, CL2D dismisses the clusters whose
size is below a certain number (30 in our case) and splits the
largest cluster into two clusters once a dismiss is executed.
Another difference is that CL2D adopts the correntropy [a~kernel-based entropy measure, \citet{Sorzano}] as the measure of
distance, while $k$-means uses the usual Euclidean norm. We thus
consider $k$-means$^+$, which is modified from CL2D by using the
Euclidean norm instead. Note that CL2D, $k$-means$^+$ and $k$-means
require the prespecification of $K$ and initial assignments for the
cluster centers. CL2D further needs a tuning parameter for the
kernel width. We have tried a few kernel widths and picked the best
one for CL2D. We present the best results out of 10 runs for CL2D,
$k$-means and $k$-means$^+$. In contrast to this, $\gamma$-SUP is
deterministic as long as its parameters $s$ and $\tau$ are fixed. As
has been demonstrated in the previous numerical studies that
$\gamma$-SUP is insensitive to the $s$-selection, we fix $s =
0.025$. We will supply a phase transition diagram in
Section~\ref{sec5.1} to choose $\tau$.

To evaluate the performance of each method, we report the impurity and
\mbox{c-impurity} numbers as defined below. Let \{$c_i$\} be sets of true
clusters, \{$\omega_j$\} be sets of constructed clusters, and $|\cdot|$
be the cardinality of the set. For each output cluster $\omega_j$, its
purity number is defined by $\max_i|c_i\cap\omega_j|$. The overall
purity number [\citet{Manning}] is the sum over all output clusters:
\[
\mbox{purity} = \sum_j \max_i
|c_i \cap\omega_j|.
\]
The impurity number is defined to be
\[
\mbox{impurity} = n - \mbox{purity}.
\]
Note that the purity is usually defined to be the ratio of the purity
number and the total number of images. Here we do not normalize it by
the total number for better presentation of the simulation results. The
impurity number is 0 for the perfect clustering result, but a zero
impurity number does not guarantee a perfect clustering. This number
cannot recognize mistakes made by splitting one class into two or more
clusters. We thus define c-impurity, which is analogous to the impurity
number but exchanges the roles of the true clusters and the output
clusters to derive its measure of impurity. That is, define
\[
\mbox{c-impurity} = n - \sum_i \max
_j |c_i \cap\omega_j|,
\]
which is able to pick up the mistakes by splitting a cluster into two
or more clusters. In summary, small values of the impurity and
c-impurity numbers indicate better performance of a clustering method.

\subsubsection{Clustering with perfectly-aligned images}
\label{sec5.1}

Simulation results with\break perfectly-aligned images are placed in
Table~\ref{table.perfect_eqnarray}. For the small noise level of
$\sigma
_\varepsilon= 40$, $\gamma$-SUP, CL2D and $k$-means$^+$ give perfect
clustering. For larger $\sigma_\varepsilon$ values, their clustering
accuracies slowly decay as expected, and have comparable performances.
CL2D confuses images at an impurity level of $(4,4)$ (which stands for
$\mbox{impurity} = 4$ and $\mbox{c-impurity} = 4$) when $\sigma
_\varepsilon= 60$. Quite unexpectedly, $k$-means$^+$ confuses images at
an impurity level of $(34, 33)$ when $\sigma_\varepsilon= 50$ but
performs flawlessly when $\sigma_\varepsilon= 60$. This is due to the 10
random initial centers mentioned above. The performance of $k$-means is
poor for all $\sigma_\varepsilon$ values, even when we correctly specify
$K = 128$. This reflects the shortcomings of $k$-means when the noise
level is high and when the number of clusters is large.

%
\begin{table}
\caption{Clustering result with perfect alignment images}
\label{table.perfect_eqnarray}
\begin{tabular*}{\textwidth}{@{\extracolsep{\fill}}lccccc@{}}
\hline
& \multicolumn{1}{c}{$\bolds{\gamma}$\textbf{-SUP}} &
\multicolumn{1}{c}{$\bolds{\gamma}$\textbf{-SUP}$\bolds{^{+}}$} & \multicolumn{1}{c}{\textbf{CL2D}} &
\multicolumn{1}{c}{$\bolds{k}$\textbf{-means}$\bolds{^{+}}$} &
\multicolumn{1}{c@{}}{$\bolds{k}$\textbf{-means}}\\
\hline
& \multicolumn{5}{c}{$\sigma_\varepsilon= 40$ (SNR${} = {}$0.19)} \\
impurity & \phantom{00}0 & 0 & 0 & \phantom{0}0 & 1206 \\
c-impurity & \phantom{00}0 & 0 & 0 & \phantom{0}0 & \phantom{0}425 \\[3pt]
& \multicolumn{5}{c}{$\sigma_\varepsilon= 50$ (SNR${} = {}$0.12)} \\
impurity & \phantom{0}44 & 0 & 0 & 34 & 1175 \\
c-impurity & \phantom{00}0 & 0 & 0 & 33 & \phantom{0}462 \\[3pt]
& \multicolumn{5}{c}{$\sigma_\varepsilon= 60$ (SNR${} = {}$0.08)} \\
impurity & 150 & 0 & 4 & \phantom{0}0 & 1106 \\
c-impurity & \phantom{00}0 & 0 & 4 & \phantom{0}0 & \phantom{0}465 \\
\hline
\end{tabular*}
\end{table}

One can see that $\gamma$-SUP has a larger impurity number when
$\sigma
_\varepsilon= 60$, and we found that the errors made by $\gamma$-SUP
always involve mistakenly combining two clusters as a single one. In
real applications, practitioners are very likely to have prior
knowledge about the expected cluster size when analyzing cryo-EM
images, and it is common to further bisect an unevenly large cluster.
To mimic this situation, we consider $\gamma$-SUP$^+$, which modifies
the result of $\gamma$-SUP by further using $k$-means to separate those
clusters whose size is greater than 70. This threshold should be
adjusted according to the ratio of the total number of images to the
number of clusters expected. The results of $\gamma$-SUP$^+$ are also
provided in Table~\ref{table.perfect_eqnarray}, where a perfect clustering
is achieved for every $\sigma_\varepsilon$. This indicates the
applicability of our proposal, as an error made by $\gamma$-SUP can be
easily fixed. We remind the readers that the true number of components
in this simulation is $K = 128 = 2^7$. This makes CL2D (and
$k$-means$^+$) beneficial in clustering, since the main idea of CL2D is
to bisect data until a prespecified number of clusters is reached. We
thus believe using $\gamma$-SUP$^+$ is a fair comparison for our
method. We will further see the superiority of $\gamma$-SUP in the
presence of outliers (see Section~\ref{sec.misasslined}), in which the
true number of components largely exceeds $128$, resulting in CL2D
becoming less accurate at clustering.

\begin{figure}[t]
\includegraphics{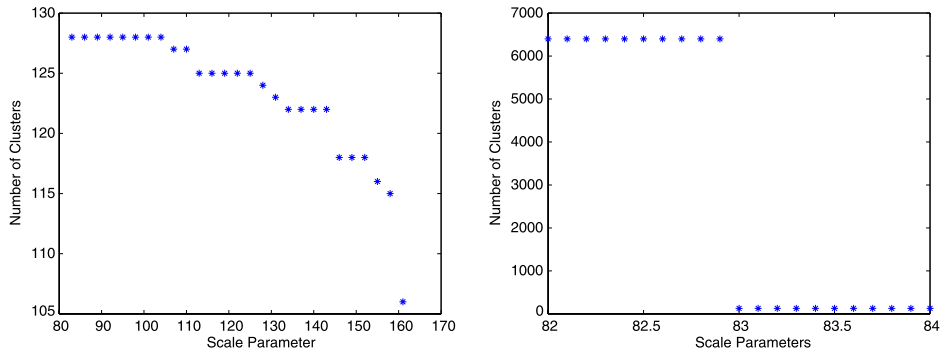}
\caption{The number of clusters created by $\gamma$-SUP under various
values of $\tau$. A phase transition occurs when the scale parameter
$\tau$
is 83.}
\label{ClassNumberScale}
\end{figure}

We demonstrate the effect of $\tau$ and provide guidance on its
selection. Figure~\ref{ClassNumberScale} gives the numbers of clusters
from $\gamma$-SUP under various values of $\tau$ when $\sigma
_\varepsilon
= 40$, wherein we observe a phase transition in the number of clusters:
$\gamma$-SUP outputs 6400 clusters when $\tau< 83$, outputs 128
clusters (a perfect result) when $\tau= 83$, and there exists no
intermediate result between 128 and 6400. Moreover, the cluster number
remains at 128 for quite a wide range of $\tau\in[83, 105]$. Recall
that the scale parameter $\tau$ is proportional to the support region
of the weight \eqref{weight} and, hence, $\gamma$-SUP's updating
procedure ignores the influence of data outside a certain range
determined by $\tau$. When $\tau$ is small enough that there is no
influence between any two images, $\gamma$-SUP leads to 6400 clusters
(i.e., each individual cryo-EM image forms one cluster). When $\tau$
reaches a critical value, the images in the same cluster can start
attracting each other and will finally merge. This explains why a phase
transition occurs. We observe similar phase transition phenomena for
various noise structures, of which some may not happen at the perfect
cluster result, but never happen far from it. Thus, the value at which
the phase transition occurs can be treated as a starting value for
selecting a reasonable range of $\tau$.

\subsubsection{Clustering with misaligned images}
\label{sec.misasslined}

It is well possible that a set of cryo-EM images cannot be well
aligned due to their low SNR. A good cluster method should be robust in
the presence of misaligned images (outliers). We thus conduct
simulations to evaluate the performances of clustering methods, where
$10\%$ and $20\%$ of the images are randomly chosen to be rotated by
7.2, 14.4, 21.6, 28.8, 36 or 43.2 angular degrees ($^\circ$) clockwise. The
effect of this is that each rotated image no longer shares the same
signal pattern with the images in its original cluster, nor does it
share a signal pattern with the other misaligned ones. An ideal outcome
in this scenario would be for the clustering algorithm to treat each of
these misaligned images as a singleton cluster. Including these
singleton clusters, the total cluster number would become 771 for $10\%
$ misalignment and 1410 for $20\%$, while the meaningful cluster
number would remain at 128.

Simulation results are presented in Table~\ref{table.10-mis_eqnarray}
for the $10\%$ misalignment scenario and
Table~\ref{table.20-mis_eqnarray} for the $20\%$ misalignment scenario.
It can be seen that $\gamma$-SUP performs best in
comparison with CL2D, $k$-means$^+$ and $k$-means for all
$\sigma_\varepsilon$ values. Moreover, combined with the prior
knowledge about expected cluster size, the result from $\gamma$-SUP
can be further improved by $\gamma$-SUP$^+$. These outcomes support
the applicability of $\gamma$-SUP in the presence of outliers. As
shown in Table~\ref{table.10-mis_eqnarray} for the very low SNR case
with $\sigma_\varepsilon= 60$, $\gamma$-SUP$^+$ has only 7
misaligned images which have been wrongly assigned to true clusters,
while the other misaligned images were correctly separated out as
outliers. On the other hand, $k$-means$^+$ and CL2D can do nothing
about the misaligned images which are $643$ ($= 771 - 128$) in the
$10\%$ misalignment case, such that their default mistakes are 643
in the impurity category. The main reason is that CL2D will still
allocate outliers into certain clusters and, hence, the resulting
cluster centers will still be subject to the influence of outliers.
In contrast, $\gamma$-SUP allows for singleton clusters as
indicated in Table~\ref{table.10-mis_eqnarray}, and the resulting
cluster means are, therefore, more robust.

\begin{table}[t]
\caption{Clustering result with $10\%$ misalignment images}
\label{table.10-mis_eqnarray}
\begin{tabular*}{\textwidth}{@{\extracolsep{\fill}}lccccc@{}}
\hline
& \multicolumn{1}{c}{$\bolds{\gamma}$\textbf{-SUP}} &
\multicolumn{1}{c}{$\bolds{\gamma}$\textbf{-SUP}$\bolds{^{+}}$} & \multicolumn{1}{c}{\textbf{CL2D}} &
\multicolumn{1}{c}{$\bolds{k}$\textbf{-means}$\bolds{^{+}}$} &
\multicolumn{1}{c@{}}{$\bolds{k}$\textbf{-means}}\\
\hline
& \multicolumn{5}{c}{$\sigma_\varepsilon= 40$ (SNR${} = {}$0.19)} \\
impurity & \phantom{00}0 & 0 & 643 & 789 & \phantom{0}407 \\
c-impurity & \phantom{00}0 & 0 & \phantom{00}0 & \phantom{00}0 & 3452 \\[3pt]
& \multicolumn{5}{c}{$\sigma_\varepsilon= 50$ (SNR${} = {}$0.12)} \\
impurity & \phantom{0}83 & 0 & 644 & 788 & 410 \\
c-impurity & \phantom{00}0 & 0 & \phantom{00}1 & \phantom{00}4 & 3482 \\[3pt]
& \multicolumn{5}{c}{$\sigma_\varepsilon= 60$ (SNR${} = {}$0.08)} \\
impurity & 190 & 7 & 644 & 779 & \phantom{0}423 \\
c-impurity & \phantom{00}0 & 0 & \phantom{00}1 & \phantom{00}2 & 3541 \\
\hline
\end{tabular*}
\end{table}
%
\begin{table}[t]
\caption{Clustering result with $20\%$ misalignment images}
\label{table.20-mis_eqnarray}
\begin{tabular*}{\textwidth}{@{\extracolsep{\fill}}lccccc@{}}
\hline
& \multicolumn{1}{c}{$\bolds{\gamma}$\textbf{-SUP}} &
\multicolumn{1}{c}{$\bolds{\gamma}$\textbf{-SUP}$\bolds{^{+}}$} & \multicolumn{1}{c}{\textbf{CL2D}} &
\multicolumn{1}{c}{$\bolds{k}$\textbf{-means}$\bolds{^{+}}$} &
\multicolumn{1}{c@{}}{$\bolds{k}$\textbf{-means}}\\
\hline
& \multicolumn{5}{c}{$\sigma_\varepsilon= 40$ (SNR${} = {}$0.19)} \\
impurity & \phantom{00}0 & \phantom{0}0 & 1703 & 1773 & \phantom{0}820 \\
c-impurity & \phantom{00}0 & \phantom{0}0 & \phantom{000}4 & \phantom{000}3 & 3883 \\[3pt]
& \multicolumn{5}{c}{$\sigma_\varepsilon= 50$ (SNR${} = {}$0.12)} \\
impurity & \phantom{0}36 & \phantom{0}1 & 1743 & 1760 & \phantom{0}833 \\
c-impurity & \phantom{00}0 & \phantom{0}0 & \phantom{000}1 & \phantom{00}11 & 3899 \\[3pt]
& \multicolumn{5}{c}{$\sigma_\varepsilon= 60$ (SNR${} = {}$0.08)} \\
impurity & 214 & 11 & 1726 & 1713 & \phantom{0}824 \\
c-impurity & \phantom{00}0 & \phantom{0}0 & \phantom{000}6 & \phantom{000}1 & 3909 \\
\hline
\end{tabular*}
\end{table}

The performance of CL2D and $k$-means$^+$ is more seriously
impacted when the misaligned proportion is $20\%$
(Table~\ref{table.20-mis_eqnarray}). While CL2D makes no other mistake
at $\sigma_\varepsilon= 40$ and makes $(1, 1)$ mistakes at
$\sigma_\varepsilon= 50$ and $\sigma_\varepsilon= 60$ in addition
to the
default 643 misalignment images in the $10\%$ case, it makes from
421 to 444 more image merges in addition to the default 1282 ($=
1410 - 128$) ones for the $20\%$ case. The $k$-means$^+$ method
shows similar patterns in both the $10\%$ and the $20\%$
case. This indicates that, as a large number of outliers are forced
to enter the clusters, their cluster representatives tend to be
contaminated to an extent that it becomes difficult to identify
their correct cluster members. In contrast, $\gamma$-SUP provides a
solution to this commonly encountered situation.

%
\begin{figure}[t]
\includegraphics{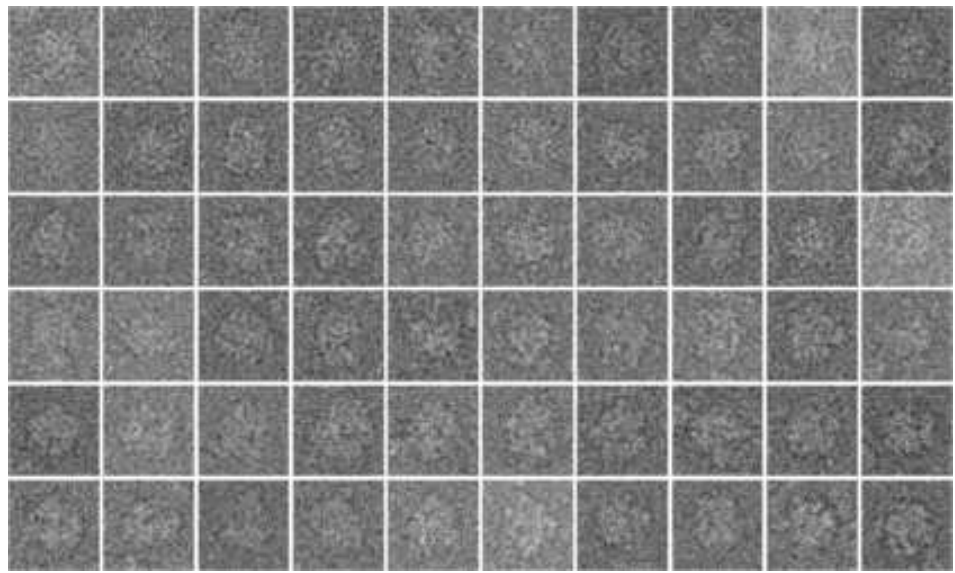}
\caption{60 images are randomly chosen from the 5000 ribosome cryo-EM
image set.}
\label{ribosomeImages}
\end{figure}

\section{Real cryo-EM image clustering}
\label{sec.real}

\begin{figure}
\includegraphics{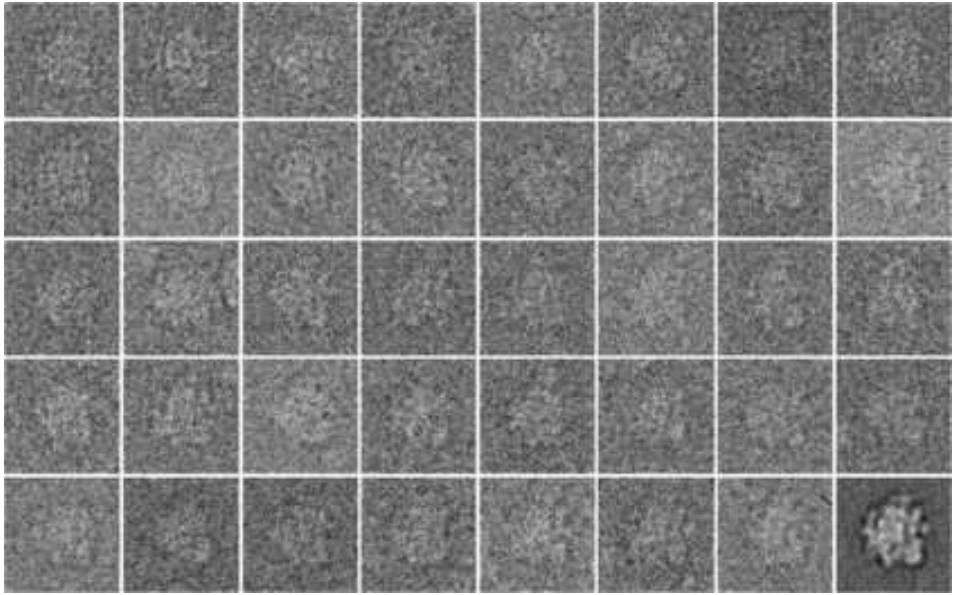}
\caption{One $\gamma$-SUP clustering class example contains 39 element
images. Their class average is shown on the right-bottom corner.}
\label{ribosomeClass}
\end{figure}

\begin{figure}
\includegraphics{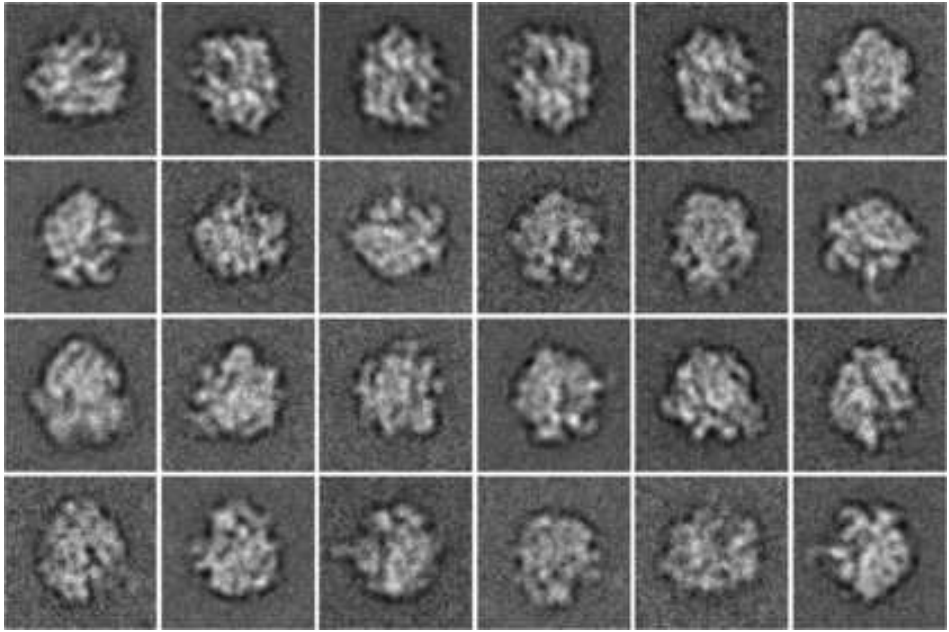}
\caption{24 class averages by $\gamma$-SUP.}
\label{classAverage}
\end{figure}

\begin{figure}[t]
\includegraphics{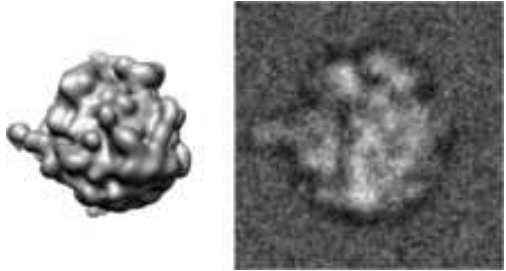}
\caption{A single projected view of a known ribosome 3D structure
compared with a single class average obtained by $\gamma$-SUP.}
\label{ribosomeCompare}
\end{figure}

Ribosome is the cellular machinery that synthesizes proteins. The
structure of ribosome has been intensively investigated by X-ray
crystallography and cryo-EM. For the latter, the structure is obtained
by 3D reconstruction using many cryo-EM images. Those images represent
projections from different viewing angles. To test the performance of
$\gamma$-SUP in grouping experimental data into similar views, a set of
E. coli 70s ribosome was downloaded\footnote{The website link is
\url{http://www.ebi.ac.uk/pdbe/emdb/data/SPIDER\_FRANK\_data/}.} (60
sample images are shown in Figure~\ref{ribosomeImages}). This data set
contains 5000 single-particle cryo-EM images of randomly oriented
ribosome carrying an elongation factor with it. The 5000 cryo-EM
images were preprocessed using XMIPP [\citet{XMIPP}] to correct the
negative phase component induced by the electron microscope transfer
function. Next, the corrected images were centered and roughly aligned
using the multi-reference alignment method on the SPIDER suite for
single particle processing [\citet{SPIDER}]. The dimension of those
images was reduced from $130\times130$ to $15\times15$ via MPCA
[\citet
{Hung}]. PCA was then further applied to the correlation matrix of
those MPCA loading scores. The final dimension of each image is 20.
Figure~\ref{ribosomeClass} shows a gallery of 39 raw cryo-EM images
with a noticeably similar orientation, which were grouped into a single
cluster by $\gamma$-SUP using the parameters $s = 0.025$ and $\tau=
1$, and the average of those images at the right bottom corner, which
has clearly brought out an abundance of meaningful detail hidden in the
raw images. This particular clustering exemplifies the potential of
$\gamma$-SUP in processing real data. A total of 24 clusters with
cluster sizes between 23 to 219 were found and their averages are given
in Figure~\ref{classAverage}. We demonstrate their consistency with the
2D projections of known ribosome 3D structure with an example shown in
Figure~\ref{ribosomeCompare}. One of the class averages is compared
with a 2D projection of a 70s ribosome 3D structure solved by X-ray.
The P-stalk [\citet{Wilsome}] with its signature tail on the left side
can be clearly seen.

%

%

%

\section{Conclusion}
\label{sec.conclusion}

We have combined a model-based $q$-Gaussian mixture clustering method,
the $\gamma$-estimator and a self-updating process, SUP, to propose
$\gamma$-SUP, a novel hybrid designed to meet the image clustering
challenges encountered in cryo-EM analysis. Characteristically, sets of
cryo-EM images have low SNR, many of which are misaligned and should be
treated as outliers, and which form a large number of clusters due to
their free orientations. Because of its capability to identify
outliers, $\gamma$-SUP can separate out the misaligned images and
create the possibility for further correcting them. Thus, we have been
able to present a successful application of $\gamma$-SUP on cryo-EM images.

Eliminating the need to set initial random cluster centers and
cluster number, $\gamma$-SUP requires only the specification of $(s,
\tau)$. We have shown that $\gamma$-SUP has robust performance over
$s$ in many scenarios. Once $s$ is chosen, the phase transition
scheme may suggest a reasonable range for $\tau$. This insensitivity
with respect to $s$-selection and the observation of the phase
transition greatly reduces the difficulty in selecting the tuning
parameters. We summarize the characteristics which make for the
success of $\gamma$-SUP:
\begin{itemize}
\item
$\gamma$-SUP adopts a $q$-Gaussian mixture model with $q < 1$, which
has compact support. Hence, it sets a finite influence range for each
component and completely rejects data outside this range.
When a data point is outside a certain cluster representative's
influence range, it contributes zero weight toward this representative.

\item
$\gamma$-SUP estimates the model parameters by minimizing
$\gamma$-divergence. The minimum $\gamma$-divergence downweights the
influence for data that deviates far from the cluster centers, which
enhances the clustering robustness.

\item
$\gamma$-SUP extracts clusters without the need of specifying the
number of components $K$ and random initial centers. It starts with
each individual data point as a singleton cluster [i.e., with a mixture
of $n$ components, see \eqref{initial}], and $K$ is data-driven.

\item
$\gamma$-SUP allows singleton or extremely small-sized clusters to
accommodate potential outliers.

\item
$\gamma$-SUP uses $\hat F^{(\ell)}$ to shrink the fitted mixture model
toward cluster centers in each iteration. Such a shrinkage acts as if
the \textit{effective temperature} is iteratively decreasing, so that
it improves the efficiency of mixture estimation.
\end{itemize}

Finally, we also remind the readers that the strength of $\gamma$-SUP
is for cases where the number of clusters is large, or data are
contaminated with noises/outliers, or cluster sizes are not balanced
[some clusters are much bigger, or smaller, than others]. However,
there is no advantage of using $\gamma$-SUP if the clustering problem
arises from normal mixtures whose components have approximately similar sizes.

\section*{Acknowledgments}
The authors gratefully acknowledge the Editor, the Associate Editor and two
reviewers for their comments, which have substantially improved this work.
The authors also acknowledge the support from the National Science
Council, Taiwan.

\printaddresses

\end{document}